\documentclass[%
amsmath,amssymb,
aps,
pre,
twocolumn,
superscriptaddress
]{revtex4-1}

\usepackage{amsmath,amsfonts}
\usepackage{ae,aecompl}   
\usepackage{soul}
\usepackage{graphics}
\usepackage{graphicx}
\usepackage{dcolumn}
\usepackage{bm}
\usepackage{systeme}

\usepackage{geometry}
\usepackage{url}
\usepackage{subfigure}
\usepackage[english]{babel}
\usepackage[utf8x]{inputenc}
\usepackage[T1]{fontenc}

\usepackage[usenames,dvipsnames]{xcolor}

\usepackage{color}
\definecolor{darkgreen}{rgb}{0,0.6,0}
\definecolor{darkblue}{rgb}{0,0,0.6}
\definecolor{darkred}{rgb}{0.6,0,0}
\definecolor{darkpurple}{rgb}{0.5,0,0.5}

\newcommand{\argc}[1]{\left[#1\right]}

\usepackage{hyperref}

\hypersetup{
	bookmarksopen=true,
	pdftitle=,
	pdfauthor=Nirvana Caballero, 
	pdftoolbar=false,           
	pdfstartview={FitH},		
	pdfmenubar=true,			
	pdfhighlight=/O,			
	colorlinks=true,			
	urlcolor=darkblue,
	citecolor=darkblue,		    
	linkcolor=darkpurple	    
}

\begin{document}

\preprint{APS/123-QED}

\title{Roughness density}

\author{Nirvana Caballero}
\email[Corresponding author: ]{Nirvana.Caballero@unige.ch}
\affiliation{Department of Quantum Matter Physics, University of Geneva, 24 Quai Ernest-Ansermet, CH-1211 Geneva, Switzerland}

\author{Thierry Giamarchi}
\affiliation{Department of Quantum Matter Physics, University of Geneva, 24 Quai Ernest-Ansermet, CH-1211 Geneva, Switzerland}

\date{\today}
\begin{abstract}
The theory of disordered elastic systems is one of the most powerful frameworks to assess the physics of multiple systems that span from ferromagnets to migrating biological cells. In this formalism, one assumes that the system can be described with a displacement field. This field can represent an interface position, the deformation of a vortex lattice or charge density waves in semiconductor devices, among others. By construction, this field is univalued and 'smooth', and, even if experimental realisations of it can be far from this description, the consequences of these approximations have not been yet fully explored. We present a new observable to measure the roughness of displacement fields that can be beyond the elastic limit and can contain overhangs and other defects. Our observable represents a stepping stone towards the construction of a general theory for interfaces. 
\end{abstract}

\maketitle

\section{Introduction}
Interfaces are collective structures that emerge as a consequence of the coexistence of two (or more) states in a system. These emergent structures minimise their energy by adopting the shortest possible length, thus becoming essentially flat for ideal clean (non-disordered) systems with two coexisting phases. In presence of thermal activation or structural disorder interfaces can, however, adopt more complicated shapes with a 'jiggling' shape: interfaces are deformed and become eventually rough. The jiggling structure is a consequence of the competition between the micro and macroscopic interactions present in the system and thus reveals its underlying physics.

One of the most successful frameworks to characterise the physics of interfaces is called \textit{Disordered elastic systems} or \textit{Disordered elastic manifolds}. In this formalism, one assumes that interfaces can be described by a displacement field $u({x})$. Beyond describing an interface position, $u$ can also represent a displacement field measuring the deformation of a vortex lattice or charge density waves in semiconductor devices. Characterising its statistical properties is thus highly relevant in many contexts~\cite{kay_review_2022}. 

The roughness of an interface can be measured as $\langle[u({x})-u({y})]^2 \rangle$, the averaged correlations of the quadratic displacements of the function $u$, and has served as a useful tool to assess the physics of different systems at multiple scales. These scales range from experimentally inaccessible distances to multiple decades, thus providing an overall assessment of a system's properties. The roughness power-law behaviour(s) reveals the (multi)-interactions dominating the physics of a system and thus, it has been widely computed to characterise diverse experimental systems with very different microscopical details that span from thin ferromagnetic~\cite{lemerle_domainwall_creep,ferre_2013_ComptesRendusPhys14_651,durin_2016_PRL_magneticavalanches,caballero2017excess,pardo2017,CortesBurgos_PhysRevB.104.144202_2021,torres2019universal,lee2009roughness,caballero2017excess,magni2009visualization,ferre2013universal,grassi2018intermittent,domenichini2019transient,Jordan2020} or ferroelectric films~\cite{paruch2013nanoscale,salje_2019_PRM_ferroelectricavalanches,tuckmantel_2021_local,cherifi2017non,ziegler2013domain,paruch2013nanoscale,guyonnet2012multiscaling,torres2019universal,lee2009roughness,magni2009visualization}, cell fronts~\cite{huergo2010morphology,muzzio2014influence,tuckmantel2021local,rapin2021roughness}, bacterial colonies~\cite{Bonachela_JSP_2011}, crack fronts~\cite{schmittbuhl_PRL_1997DirectObservationOfASelfAffineCrackPropagation,delaplace_PRE_1999HighResolutionDescriptionCrackFrontPlexiglasBlock} to contact lines~\cite{moulinet2004width,leDoussal2009heightfluctuationsContactLine}.

The roughness scaling, in conjunction with that of other observables (as for example, the scaling of the overall interface velocity) has been successfully used to determine the universality class to which a numerical, experimental or analytical system belongs~\cite{laurson_JSTAT_2010RoughnessAndMultiscalinPlanarCrackFronts,Ferrero2013,domenichini2019transient,caballero_JSTAT_2021_ac,rapin2021dynamic,ferrero_2021_AnnualReviewsCondMattPhys_creep,caballero_PRE_2022_09}. The available tools to compute the roughness of an interface rely on the determination of the function describing the interface position $u$~\cite{giamarchi_2006_arXiv:0503437,kay_review_2022}. This definition assumes, by construction, that the interface is univalued and lacks an internal structure (for example a finite width). It is, however, highly non-trivial in some cases to determine the function $u$, and in many cases, one has to depend on uncontrolled approximations. Besides, the effects of overhangs are essential to understand the overall behaviour of interfaces~\cite{grassberger_2020_PRR_sirpercolation}.

In this work, we present a new observable to assess the roughness of interfaces. Our observable relies on the computation of the correlations of the Fourier transform of the interface 'density', thus being readily applicable to the study of a broad spectrum of interfaces, including those multivalued or with an internal structure. Our observable thus represents a stepping stone towards the construction of a general theory for interfaces, since it allows for the characterisation of the effects of structures that are usually discarded due to the lack of proper tools to characterise their properties.

We showcase the potential of our observable by analysing interfaces in a Ginzburg-Landau model. This allows us to benchmark our observable: we show how our observable can capture the roughness of interfaces subject to thermal fluctuations and recover analytical predictions.

Our tool allows us to probe the roughness of systems at high temperatures, where it is seemingly more difficult to determine an univalued function describing the interface position. We show that even in this case, the roughness can be very well described by a power-law with the expected thermal exponent $\zeta=0.5$, but with a prefactor higher than the one predicted by the elastic theory. Moreover, we show how our procedure for interface detection, based on the zero-temperature solitonic solution is an accurate method to detect interfaces in these models~\cite{caballero_prb_2020_qEW-GL}.

Our observable opens the path to explore the effects of defects in interfaces' roughness, the effects of different approximations usually used to force interfaces to an elastic line description and moreover, it can allow others to characterise highly irregular experimental realisations of interfaces. 

\begin{figure*}[bp!]
\begin{center}
 \includegraphics[width=1\linewidth]{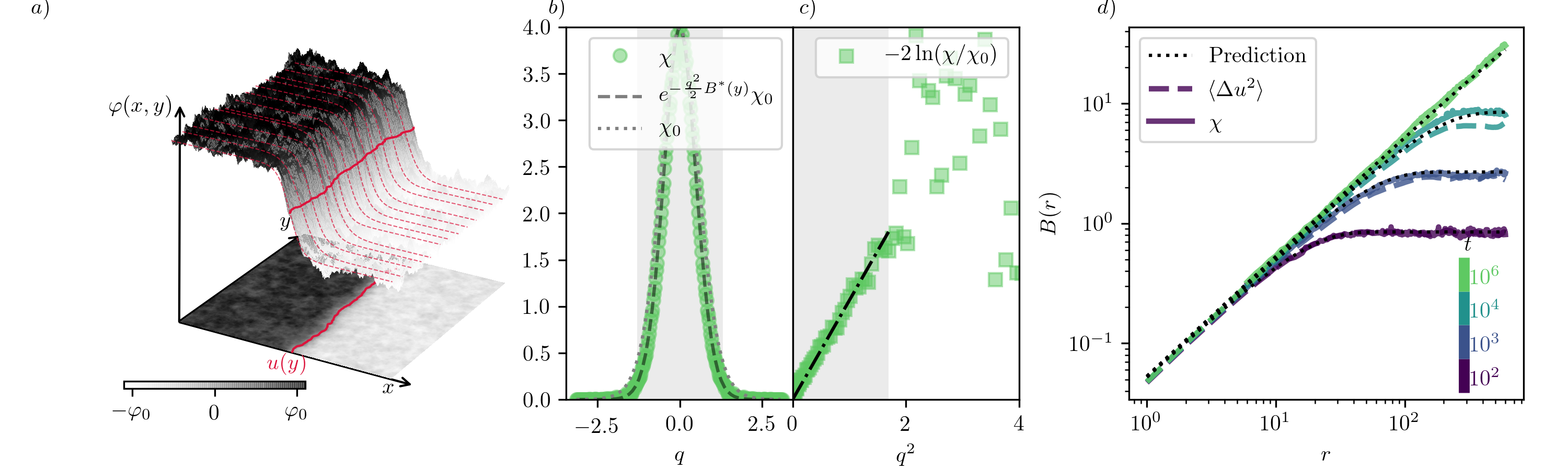}
\end{center}
\caption{a) Part of an interface in a Ginzburg-Landau model that evolved for a time $t=10^6$ from a flat initial condition at $T=0.05$. Some of the solitonic profiles used to fit $\varphi(x,y)$ along the $x$-direction to obtain the approximate interface position $u(y)$ (continuous red line) are shown in red dashed lines. b) Roughness density $\chi$ (as defined in Eq.~\ref{eq:GLRoughnessDensityDefinition}) for $y=20$ as a function of $q$, the Fourier transform along the $x$-direction variable. For comparison, we also show the quantity $e^{-q/2B^*(y=20)}\chi_0$ (dashed lines), where $B^*(y)$ is the roughness of the interface $u(y)$ computed as $\langle \Delta u^2 \rangle$: this quantity is in excellent agreement with $\chi$. The form factor $\chi_0$ (Eq.~\ref{eq:chi0GL}) is shown in dotted lines. c) To obtain the roughness from $\chi$, we compute $-2\ln(\chi/\chi_0)$ (here shown in squares for $y=20$) and fit it with a linear function of $q^2$ in the region that gives the best fit, highlighted in grey (also shown in b)). The slope of the fitted function is the value of the roughness obtained from $\chi$. d) Roughness of a Ginzburg-Landau interface that evolved from a flat initial condition evaluated at different simulation times $t=10^i$, for $i=2,3,4,6$ averaged over 10 realizations obtained with two different methods: one through the detection of the interface position $u(y)$ by fitting the order parameter $\varphi$ with a solitonic solution (dashed lines), and a second method presented in this work that relies on the computation of the roughness density $\chi$ (Eq.~\ref{eq:RoughnessDensityDefinition}) (continuous lines). The roughness of this interface behaves as that of an elastic line whose dynamic evolution is ruled by an Edwards-Wilkinson equation (dotted black lines, given by Eq.~\ref{eq:B(r,t)fromFlat}) with equivalent parameters.
}
\label{fig:B_chi_time}
\end{figure*}

\section{Observable definition}

Given an interface described by a function $u(y)$ $ \mathbb{R}\rightarrow\mathbb{R}$ separating two different homogeneous states of a system hosted by a surface $\Omega \in \mathbb{R}^2$ we can define the interface density as 
\begin{equation}
\rho(x,y)=\delta(x-u(y))=\int_\Omega d\lambda e^{i\lambda(x-u(y))}.
\end{equation}

Its one-dimensional Fourier transform in $x$ satisfies $\mathcal{F}[\rho(x,y)]=\tilde \rho(q,y)=\int_{B_\Omega} \delta(x-u(y))e^{-iqx}dx=e^{-iqu(y)}$, so we can write
\begin{equation}
 \begin{aligned}
\chi(q,y) &  =\langle \tilde \rho(q,y) \tilde \rho(-q,y') \rangle= e^{-\frac{q^2}{2}\langle [u(y')-u(y)]^2 \rangle}\\
& = e^{-\frac{q^2}{2}B(r)},
\end{aligned}
\label{eq:RoughnessDensityDefinition}
\end{equation}
where we have assumed that the stochastic variable $u$ obeys a Gaussian distribution and the roughness is $B(r=y'-y)=\langle [u(y')-u(y)]^2 \rangle$. Eq.~\ref{eq:RoughnessDensityDefinition}, shows that the observable $\chi$ can give us access to the roughness of an interface without the need to define the univalued function $u(y)$ nor the interface position, provided that a 'density' for the interface can be computed.

\section{Assessing the roughness of interfaces in a Ginzburg-Landau model}

Our observable $\chi$ (Eq.~\ref{eq:RoughnessDensityDefinition}) is a general tool to assess the roughness of interfaces defined in a surface. Here, we showcase its potential by applying it to analyse interfaces in a Ginzburg-Landau-type model. These versatile models have been proven useful to assess the effects of different protocols over interfaces in conjunction with bulk properties~\cite{caballero2018magnetic,guruciaga2021tuning,caballero_JSTAT_2021_ac,caballero_2022_arxiv_membranes}. At the same time, these models provide a benchmark to analyse the applicability of our observable. It can be shown that a Ginzburg-Landau model can be reduced to an elastic line description~\cite{caballero_prb_2020_qEW-GL} for which precise analytical calculations can be performed.

We consider a system with two preferred homogeneous states described by a local non-conserved order parameter $\varphi(\vec{r},t)$.
The system is ruled by a Hamiltonian $\mathcal{H}([\varphi])= \int  d\vec{r}\argc{\frac{\gamma}{2}|\nabla_{\vec{r}} \varphi|^2 + V(\varphi)}$, where $\vec{r}\in \mathbb{R}^2$ and $V(\varphi)= -\frac{\alpha}{2}\varphi^2+\frac{\delta}{4}\varphi^4$ is the double-well potential establishing the value of the order parameter in the two preferred homogeneous states ${\pm \varphi_0 =\pm \sqrt{\alpha/\delta}}$ and ${\gamma}$ is the amplitude of the elastic cost associated to deformations of~$\varphi$.

At finite temperature $T$ the order parameter evolution is given by a Langevin equation 
\begin{equation}
\eta \partial_t \varphi
= -\frac{\delta \mathcal{H}[\varphi]}{\delta \varphi}+\xi = \gamma\nabla_{{\vec{r}}}^2\varphi -V'(\varphi) +\xi \, , \\ 
\label{eq:Langevin}
\end{equation}
where $\eta$ is the microscopic friction and $\xi=\xi(\vec{r},t)$ is a Gaussian white noise with zero mean and two-point correlator
\begin{equation}
\langle \xi({\vec{r}}_2,t_2)\xi({\vec{r}}_1,t_1) \rangle=2 \eta T \delta^2({\vec{r}}_2-{\vec{r}}_1)\delta(t_2-t_1).
\label{eq:thermalcorrelations}
\end{equation}

The interface density can be defined through a Fourier transform of the gradient of the order parameter over $x$
\begin{equation}
 \tilde \rho(q,y)=\mathcal{F}_x[\nabla_x \varphi(x,y) ]=-iq\tilde\varphi(q,y)
\end{equation}

The roughness density (\ref{eq:RoughnessDensityDefinition}) is then given by
\begin{equation}
 \begin{aligned}
\langle \tilde \rho(q,y) \tilde \rho(-q,y') \rangle= q^2\langle \tilde\varphi(q,y) \tilde\varphi(-q,y') \rangle.  
\end{aligned}
\label{eq:GLRoughnessDensityDefinition}
\end{equation}

Since we are interested in assessing the roughness of interfaces we consider a system with Dirichlet boundary conditions $\varphi(\pm \infty)=\mp\varphi_0$. The soliton or kink-type profile that minimises the energy of the system at zero temperature in absence of disorder under these boundary conditions is given by $\varphi^*$, such that $-\frac{\delta \mathcal{H}[\varphi]}{\delta \varphi}\Big|_{\varphi^*}=0$. The explicit solution is 
\begin{equation}
 \varphi^*(x,y)=-\varphi_0 \tanh \Big(\frac{x-u(y)}{w}\Big).
 \label{eq:Soliton}
\end{equation}

The parameter $w=\sqrt{\frac{2\gamma}{\alpha}}$ represents the width of the interface, while $\varphi_0=\sqrt{\frac{\alpha}{\delta}}$ are the preferred values $\pm\varphi_0$ for the order parameter, and $u(y)$ defines the interface position.

The solitonic solution (\ref{eq:Soliton}) implies that from (\ref{eq:GLRoughnessDensityDefinition}) we can obtain the roughness as
\begin{equation}
    B(r)=-\frac{2}{q^2}\ln \frac{\chi(q,r)}{\chi_0(q)},
    \label{eq:chiGL}
\end{equation}
where we have defined the form factor
\begin{equation}
\chi_0= \frac{(w\pi q)^2}{\sinh^2 (w\pi q/2)},
\label{eq:chi0GL}    
\end{equation}
This corresponds to the roughness density of a Ginzburg-Landau interface at zero temperature. Getting the roughness of an interface only relies on knowing the function $\varphi(x,y)$ describing the state of the system, as is usually the case in experiments. For example, when observing domain walls in ferroelectrics with atomic force microscopy or ferromagnets with polar magneto-optic Kerr effect microscopy one usually obtains a two-dimensional image where each pixel (whose size is determined by the experimental resolution) represents the system's local state of the polarisation or magnetisation, respectively. In this case we can have access to the roughness by computing $\chi(q,r)$ which does not require defining a function describing the interface position. We highlight that the form factor $\chi_0(q)$ is system-dependent. If the form factor is not known for a particular system then the roughness will be obtained up to a prefactor. 

To illustrate how our observable works, we numerically solve (\ref{eq:Langevin}) in a system of size $L\times L_s$ with fixed Dirichlet boundary conditions (we fix $\varphi=\pm\varphi_0$ in the region $x\in [0,L_D]$ and $[L_s-L_D,L_s]$, with $L=4096$, $L_s=256$, and $L_D=8$) along the $x$-direction and periodic boundary conditions in $y$. Since a linear transformation of (\ref{eq:Langevin}) allows us to write the equation in reduced units, in the following, distance time and temperature are given in units of $\sqrt{\frac{\gamma}{\eta}}$, $\frac{\eta}{\alpha}$ and $\frac{\sqrt{\gamma \eta}}{\alpha}$, respectively. We use a discretization step in time equal to $0.1$ and in space equal to $1$. 

Contrary to the elastic line description where an interface position is explicitly described, in a Ginzburg-Landau model the interface of a system is given by a finite-size region where the order parameter continuously changes from one preferential value to the other ($\pm\varphi_0$). To benchmark the roughness extraction through our observable $\chi$ (Eq.~\ref{eq:RoughnessDensityDefinition}) we first 'detect' the interface position by assuming that it can be defined by a one-dimensional function $u(y)$. We fit $\varphi(x,y)$ at each $y$-value with the solitonic profile (Eq.~\ref{eq:Soliton}) with fitting parameters $\hat \varphi_0$, $\hat u$ and $\hat w$, which is the exact solution at zero temperature and a precise method at sufficiently small temperatures, as discussed in the following.

An interface described by a function $u_{EW}(y,t)$ evolving according to the Edwards-Wilkinson equation~\cite{edwards_wilkinson}
\begin{equation}
\tilde \eta \partial_t u_{EW}= c\partial^2_y u_{EW} +\tilde \xi, 
\label{eq:EW}
\end{equation}
with friction $\tilde \eta$, elasticity $c$, and temperature~$T$~\cite{caballero_prb_2020_qEW-GL} which was initially flat has a roughness evolution given by
\begin{equation}
\begin{aligned}
B(r,t)=& \frac{Tr}{c} \bigg[1- \frac{1}{\sqrt{\pi}zr}\Big(e^{-z^2r^2}-1\Big) - \frac{2}{\sqrt{\pi}}\int_0^{zr}e^{-t^2}dt \bigg],\\
\label{eq:B(r,t)fromFlat}
\end{aligned}
\end{equation}
where $z=\sqrt{\frac{\tilde \eta}{8ct}}$. One can show that by relating the elastic line and Ginzburg-Landau parameters as $\tilde \eta=\eta\frac{2\sqrt{2}}{3}\frac{\alpha}{\delta}\sqrt{\frac{\alpha}{\gamma}}$ $c=\frac{2\sqrt{2}}{3}\frac{\alpha}{\delta}\sqrt{\alpha\gamma}$~\cite{caballero_prb_2020_qEW-GL} an interface evolving at sufficiently small temperatures will show the same roughness evolution at both levels of description.

As can be observed in Fig.~\ref{fig:B_chi_time} our method for interface detection with the solitonic fitting procedure is very precise in the small-fluctuations limit (and fairly good at higher temperatures as we discuss later in the manuscript): the roughness of interfaces in a clean system and relatively low temperature ($T=0.05$) behave as the expected one for equivalent elastic lines. In this case, we obtain a Gaussian probability distribution for $\hat w$ around the equilibrium value $w=\sqrt{2}$ with standard deviation $\sigma_{\hat w}=0.25$, and $\hat \varphi_0$ as well, Gaussian centred in $0.99$, with standard deviation $\sigma_{\hat \varphi_0}0.01$ at all the studied times. 

We obtain the roughness through our observable with (\ref{eq:chiGL}). For each simulation we compute $\chi(q,r)$ (Eq.~\ref{eq:GLRoughnessDensityDefinition}). Since the solitonic solution is only exact at zero temperature, our observable will reflect all the fluctuations present in the system, including those at the bulk level. The bulk fluctuations are only reflected in our observable at short distances ($r<5$, see SM for a discussion on the bulk contribution to the roughness density). We obtain the roughness $B(r)$ by fitting $f(q)=-2\ln \frac{\chi(q,r)}{\chi_0(q)}$ with a linear function of $q^2$ (by assuming a zero intercept): the slope of the fitted function is the roughness at a distance $r$ (Eq.~\ref{eq:chiGL}). In Fig.~\ref{fig:B_chi_time} we show how this method gives access to the roughness function of an interface without having to define an univalued function describing the interface position. By adding disorder to the system ($\varepsilon=1$) we observe that the obtained roughness is indistinguishable for both methods (with the interface detection and with our new observable $\chi$), as shown in the Supplementary Material (SM).

Our observable $\chi$ serves to assess the roughness of interfaces without the need to define the interface position. This has several advantages, for example, allowing us to analyse the roughness of interfaces which are beyond the elastic limit, i.e., those with overhangs or those for which the solitonic fitting procedure can be a rough approximation. However a deeper analysis of this situation will be done in a forthcoming paper, here, we probe the roughness of interfaces subjected to high temperatures for which fluctuations break the elastic line behaviour.

\begin{figure*}[bp!]
\begin{center}
 \includegraphics[width=1\linewidth]{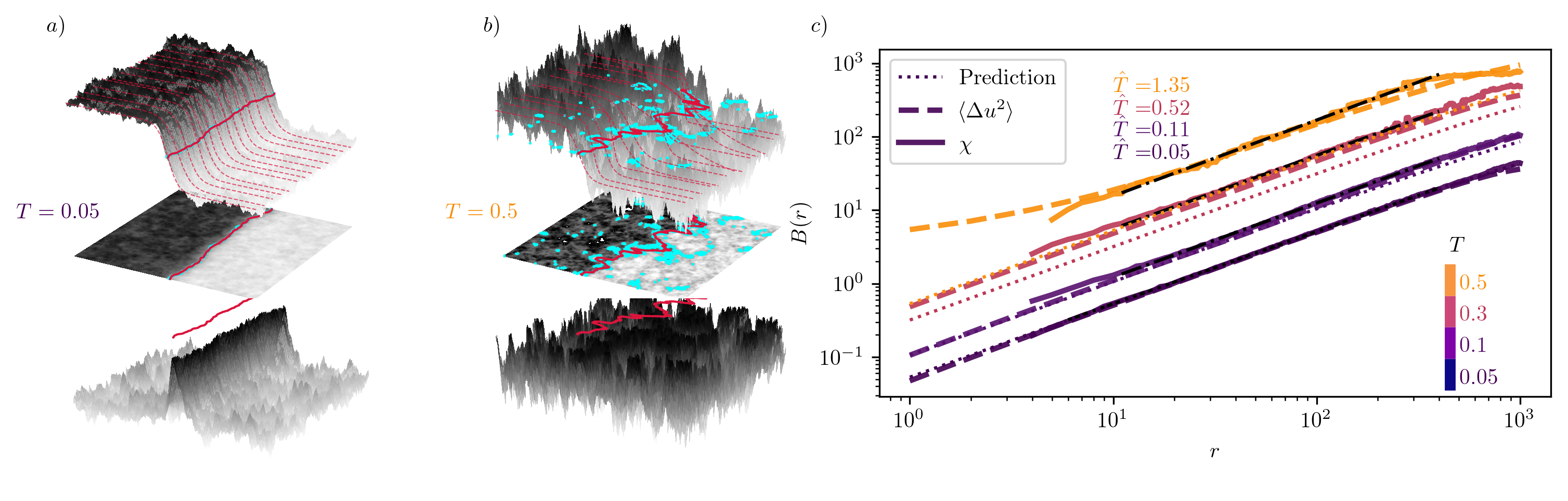}
\end{center}
\caption{Configurations of the Ginzburg-Landau order parameter $\varphi(x,y)$ at a) the lowest ($T=0.05$) and b) highest ($T=0.5$) studied temperatures after evolving for a time $t=10^6$ from a flat initial condition and its projections on the $x-y$ plane (top) and minus their absolute value (bottom). We show in red dashed lines some of the solitonic profiles resulting from fitting $\varphi(x,y)$ at each $y$-value to determine the interface position. For comparison, and to highlight the irregular nature of the interface at the higher temperature, we show in cyan all those contours where $\varphi$ is equal to zero. At the lowest temperature, both methods of interface detection give very similar results. However, this is not the case at the highest temperature where the bulk and interface fluctuations are approximately of the same order. c) Roughness of Ginzburg-Landau interfaces that evolved from a flat initial condition at different temperatures $T=0.05,0.1 ,0.3,0.5$ averaged over 50 realisations (100 for the highest studied temperature) obtained with two different methods: one through the detection of the interface position $u(y)$ by fitting the order parameter $\varphi$ with a solitonic solution (dashed lines), and a second method presented in this work that relies on the computation of the roughness density $\chi$ (Eq.~\ref{eq:RoughnessDensityDefinition}) (continuous lines). At small temperatures, the roughness of these interfaces behaves as those of elastic lines whose dynamics are ruled by an Edwards-Wilkinson equation (dotted lines, given by Eq.~\ref{eq:B(r,t)fromFlat}) with equivalent parameters. The fit of the roughness obtained through our method with a power-law $\hat Tr^{2\hat\zeta}/c$ are shown in dot-dashed lines. In all cases, we obtain $\hat \zeta=0.5$, and an effective temperature as indicated in the same colour in the figure.}
\label{fig:B_chi_temperature}
\end{figure*}

We now study the roughness of interfaces that evolved for a time $t=10^6$ at different temperatures $T=0.05,0.1,0.3$ and $0.5$ with our two methods. \textit{i)} The first one is based on fitting $\varphi$ at each $y$-value with a solitonic solution from where we extract the interface position $u$ and then compute its quadratic correlations, while in the second one \textit{ii)} we use our new observable, the roughness density of eq.~\ref{eq:GLRoughnessDensityDefinition}, for which the roughness determination is reduced to find the slope of the best fit of $-2\ln (\chi/\chi_0)$ with a linear function of $q^2$ at each position $y$. As observed in Fig.~\ref{fig:B_chi_temperature} the roughness of Ginzburg-Landau interfaces that evolved from a flat initial condition at different temperatures $T=0.05,0.1 ,0.3,0.5$ averaged over 50 realisations (100 for the highest studied temperature) in all cases follows a power-law behaviour. 
At small temperatures, the roughness of these interfaces behaves as those of elastic lines whose dynamics are ruled by an Edwards-Wilkinson equation given by Eq.~\ref{eq:B(r,t)fromFlat} with equivalent parameters. Remarkably, both methods of roughness computation are equivalent at all studied temperatures. In particular, even though the interface is not univalued at the highest studied temperatures, the effect of these 'defects' is to increase the roughness prefactor while keeping the expected thermal power-law exponent. Fittings of the roughness obtained through our method with a power-law $\hat Tr^{2\hat\zeta}/c$ are shown in Fig.~\ref{fig:B_chi_temperature}. In all cases we obtain a roughness exponent $\hat \zeta=0.5$ and a higher effective temperature than the one predicted by the elastic theory. 
To highlight the irregular structure of the interfaces at the highest studied temperature, we have computed the contours for which the order parameter takes a zero value. As shown in Fig.~\ref{fig:B_chi_temperature}, the contour plot of the interface at the lowest studied temperature is almost indistinguishable from the interface position $u(y)$ obtained through the solitonic profile fitting procedure. However, at the highest studied temperature, the zero-contour detection showcases the appearance of many 'defects' as bubbles and overhangs, for which an interface is not well-defined, both methods of roughness computation give essentially the same result.

\section{Conclusions}

We present a new observable to assess the roughness of interfaces defined over a surface. Our observable relies on the calculation of the roughness density, and as a consequence does not require defining a function describing the interface position.  

We showcase the potential of our method by benchmarking its predictions with a Ginzburg-Landau model for which analytical calculations can be made in the small fluctuations limit. We show that the roughening in time of an initially flat interface as a consequence of the competition of thermal fluctuations and the interface elasticity is very well captured by our observable. Moreover, we study the roughness of highly irregular interfaces beyond the small fluctuations limit at high temperatures. Despite the highly irregular features of interfaces in this case, we show that the roughness still follows a power-law behaviour with the thermal power-law exponent predicted by the elastic theory. However, we observe that the prefactor of this power law, predicted to be essentially -in the infinite time and space limit- the coefficient of the temperature and the interface elasticity, is higher than the expected one. Our results show that highly fluctuating interfaces have a higher effective temperature.

Our observable is highly relevant in a two-fold context. From an experimental point of view, realisations of interfaces with standard imaging techniques usually reveal the system's state. In this case, our observable provides a measure of the interface roughness directly over the analysis of the experimentally obtained image (and does not require obtaining the interface position). From a theoretical point of view, our observable allow us to explore the so far intractable consequences of bubbles and overhangs in the roughness calculation, and thus in the determination of the universality class to which a system belongs. We explore these consequences in a forthcoming paper.

\begin{acknowledgments}
We thank discussions with Jean-Pierre Eckmann, Vivien Lecomte and Kay Wiese. This work was supported in part by the Swiss National Science Foundation under Division II. All numerical simulations were performed at the University of Geneva on the \textit{Mafalda} cluster of GPUs. 
\end{acknowledgments}

\typeout{}


%

\appendix
\clearpage

\onecolumngrid
\begin{center}
\medskip
\begin{large}
\textbf{\\
\textit{Supplementary material}}
\end{large}
\end{center}

\bigskip

\twocolumngrid
\renewcommand{\theequation}{S.\arabic{equation}}

\section*{Bulk contribution to the roughness density}
\label{app:bulkContribution}

At short distances $y$, the bulk fluctuations play an important role and hinder the roughness interface contribution to the observable $\chi$. The bulk contribution to our observable can be computed by proposing a fluctuating soliton solution for Eq.~\ref{eq:Langevin} of the form

\begin{equation}
 \varphi_\Lambda(x,y,t)=\varphi^*(x-u(y,t))+\Lambda(x,y,t),
 \label{eq:fluctuatingSoliton}
\end{equation}
and a fluctuating density

\begin{equation}
 \rho(x,y)=\frac{1}{w}\rho_0\Big(\frac{x-u(y)}{w}\Big) + \Gamma(x,y),
 \label{eq:FluctuatingDensity}
\end{equation}
where $\Gamma=\nabla_x \Lambda$ takes into account the thermal fluctuations around the equilibrium density $\rho_0$, and satisfy $\langle \Gamma(x,y)\Gamma(x',y')\rangle =R_\Gamma(x-x',y-y')$.

Now, our observable $\chi$ is given by

\begin{equation}
 \begin{aligned}
 \chi(q_x,y)=\langle \tilde \rho(q_x,y)\tilde \rho(-q_x,y)\rangle =\\
 \langle e^{iq_x (u(y)-u(y'))}\rangle \Big[\tilde R(q_x,w)+\tilde R_\Gamma(q_x,y'-y) \Big]
 \end{aligned}
\end{equation}
where we have defined the functions
\[
 \tilde R(q_x,w)=\int^{\infty}_{-\infty}dxdx' e^{iq_xx}\frac{1}{w^2}\rho_0\Big(\frac{x-x'}{w}\Big)\rho_0\Big(\frac{0}{w}\Big),
\]

\begin{equation}
\begin{aligned}
 \tilde R_\Gamma(q_x,w)=& \\
\int^{\infty}_{-\infty}dxdx' e^{iq_x(x-x')} \langle \Gamma(x+u(y),y)\Gamma(x'+u(y'),y')\rangle.
\end{aligned}
\end{equation}



To obtain an equation for $\Lambda$ we assume that $u$ and the bulk fluctuations are independent. In Fourier space, and assuming that eq.~\ref{eq:fluctuatingSoliton} is a solution of the Langevin equation (eq.~\ref{eq:Langevin}) we obtain 

\begin{equation}
\eta \partial_t\tilde \Lambda= [V'_0  - \gamma(q^2_x + q^2_y)]\tilde \Lambda +\tilde \xi,    
\label{eq:Lambda}
\end{equation}
where we have expanded $-V'(\varphi+\Lambda)\simeq -V'(\varphi)+\Lambda V'_0$ and used the fact that $\frac{\delta H}{\delta \varphi}|_\varphi^*=-\gamma \varphi^{*''} + V'(\varphi^*)=0$.
Eq.~\ref{eq:Lambda} has a solution $\tilde \Gamma=iq_x \tilde \Lambda=\int_0^tdt' e^{(V'_0-\gamma \vec{q}^2)(t-t')}\tilde\xi_{t'}$, whose correlations satisfy

\begin{equation}
 \langle \tilde \Gamma_{\vec{q},t} \tilde \Gamma_{-\vec{q},t} \rangle=\frac{-\eta Tq_x^2}{\gamma (A^2+q_y^2)}\Big[ e^{-2\gamma (A^2+q_y^2)t} -1\Big]
\end{equation}

with $V'_0=-2\alpha$, $A^2=q_x^2+2\alpha/\gamma$. For $t\to\infty$

\begin{equation}
\tilde R_\Gamma(q_x,y,t\to \infty)=\frac{T}{\gamma}q_x^2\frac{e^{-y\sqrt{q_x^2+2\alpha/\gamma}}}{\sqrt{q_x^2+2\alpha/\gamma}} 
\label{eq:bulkContribution}
\end{equation}



\begin{figure*}
\begin{center}
 \includegraphics[width=1\linewidth]{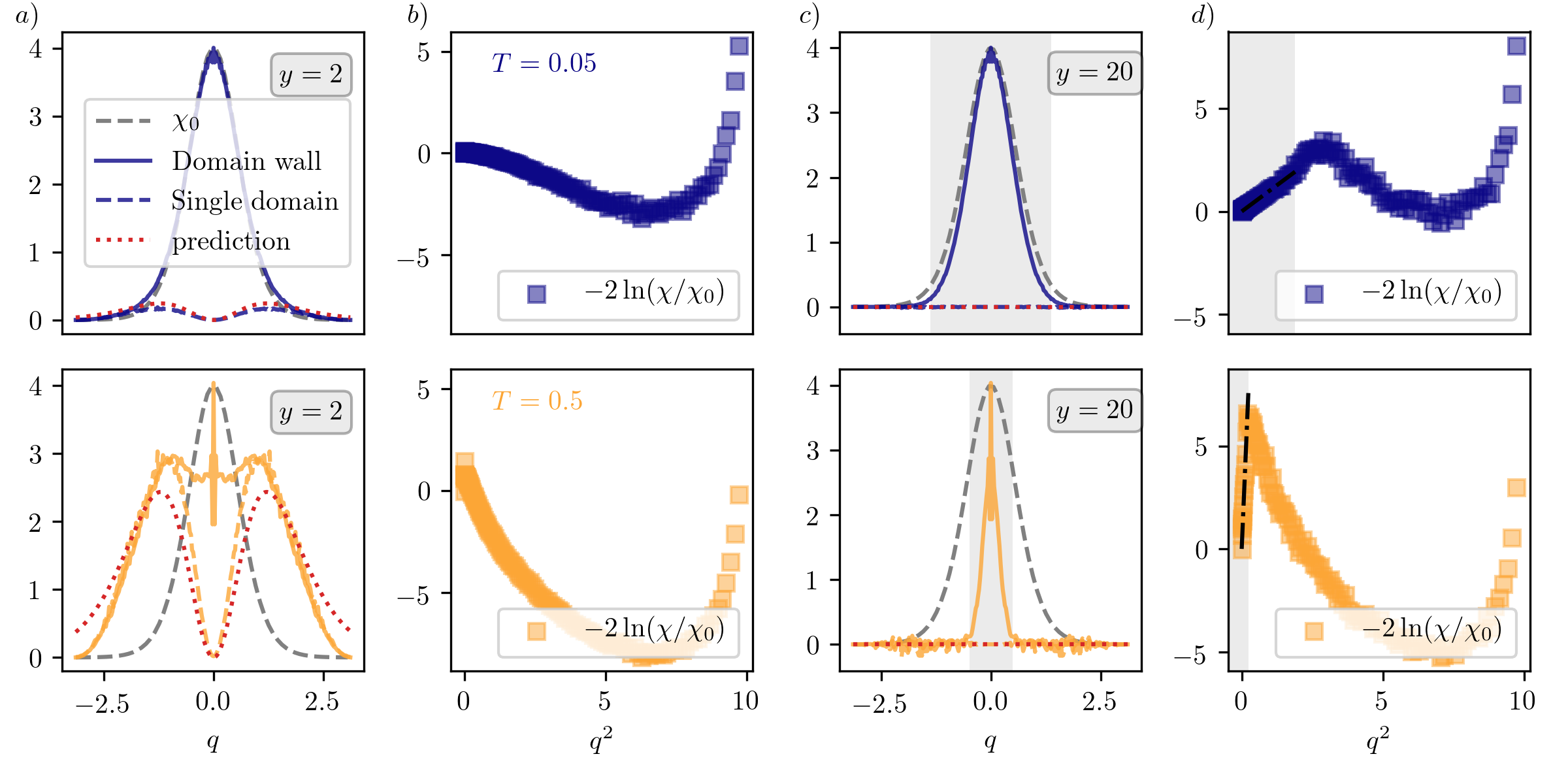}
\end{center}
\caption{We compute the roughness density $\chi$ (Eq.\ref{eq:GLRoughnessDensityDefinition}) for a domain wall (continuous lines) and a single domain (dashed lines) in a Ginzburg-Landau model at temperature $T=0.05$ (violet) and $T=0.5$ (orange) averaged over 30 realisations for a) $y=2$ and c) $y=10$. The bulk contribution can be roughly estimated by assuming a fluctuating solitonic solution (eq.~\eqref{eq:bulkContribution}) (shown in dotted red lines). The form factor corresponding to the zero temperature roughness density, $\chi_0$, is also shown in grey dashed lines. In b) and d) we show the value of $-2\ln(\chi/\chi_0)$. In c) and d) we show the region for which we get the best fit of $-2\ln(\chi/\chi_0)$ as a function of $q^2$ with a linear function.}
\label{fig:bulkContribution}
\end{figure*}

As shown in Fig.~\ref{fig:bulkContribution} this function describes fairly well the bulk contribution to our observable $\chi$ (eq.~\ref{eq:RoughnessDensityDefinition}).

\end{document}